\documentstyle[12pt]{article}
\newlength{\extraspace}
\newlength{\extraspaces}
\newcommand{\be}{\begin{equation}
\addtolength{\abovedisplayskip}{\extraspaces}
\addtolength{\belowdisplayskip}{\extraspaces}
\addtolength{\abovedisplayshortskip}{\extraspace}
\addtolength{\belowdisplayshortskip}{\extraspace}}
\newcommand{\ee}{\end{equation}}
\newcommand{\ba}{\begin{eqnarray}
\addtolength{\abovedisplayskip}{\extraspaces}
\addtolength{\belowdisplayskip}{\extraspaces}
\addtolength{\abovedisplayshortskip}{\extraspace}
\addtolength{\belowdisplayshortskip}{\extraspace}}
\newcommand{\ea}{\end{eqnarray}}
\newcommand{\nonu}{\nonumber \\[.5mm]}
\newcommand{\A}{&\!\!\!}
\newcommand{\Z}{{\bf Z}}
\newcommand{\C}{{\bf C}}
\newcommand{\R}{{\bf R}}
\newcommand{\tr}{\, {\rm tr}}
\newcommand{\e}{\, {\rm e}}

\newcommand{\bra}[1]{\left\langle {#1} \right\vert}
\newcommand{\ket}[1]{\left\vert {#1} \right\rangle}

\begin{document}
\thispagestyle{empty}
\begin{flushright}
STUPP--93--133 \\ January 1993
\end{flushright}
\vspace{.6cm}
\begin{center}
{\large{\bf{Physical States in Two-Dimensional \\[2mm]
Topological Gauge Theories}}} \\[20mm]
{\sc Yoshiaki Tanii} \\[7mm]
and \\[7mm]
{\sc Masakazu Yamashita} \\[12mm]
{\it Physics Department, Saitama University \\[2mm]
Urawa, Saitama 338, Japan} \\[20mm]
{\bf Abstract} \\[1cm]
{\parbox{13cm}{\hspace{5mm}
Physical states of two-dimensional topological gauge theories are
studied using the BRST formalism in the light-cone gauge.
All physical states are obtained for the abelian theory.
There are an infinite number of physical states with different
ghost numbers. Simple examples of physical states in a non-abelian
theory are also given.}}
\end{center}
\vfill
\newpage
\setcounter{section}{0}
\setcounter{equation}{0}
\addtolength{\baselineskip}{1mm}
%
%
%
Two-dimensional topological gauge theories have recently
appeared in various places of field theoretical models.
They have an action which can be obtained from the Chern-Simons
action in three dimensions \cite{WITTEN} by a dimensional
reduction. The SL(2, \R) topological gauge theory was
shown \cite{FK,TERAO} to describe two-dimensional gravity coupled
to a scalar field \cite{JT}.
The world-sheet action for two-dimensional stringy black-hole near
the spacetime singularity contains an abelian topological gauge
theory as a part \cite{WE}. The topological gauge theories are also
interesting by themselves as an example of topological
conformal field theories \cite{EY}.
\par
The purpose of the present paper is to investigate the structure of
physical states in the topological gauge theories.
We will use the BRST formalism in the light-cone gauge
developed for the gauge group SL(2, \R)
in Ref.\ \cite{TERAO}.\footnote{In Ref.\ \cite{TERAO} the
term `light-cone gauge' was used for a different gauge
condition from ours.}
Physical states are identified with elements of the BRST
cohomology. We mainly study the abelian theory and obtain all its
physical states.
There are an infinite number of physical states with different
ghost numbers.
The physical states are generated by the zero modes of the fields
acting on various vacua.
The non-zero modes of the fields are excluded from physical states
due to the BRST quartet mechanism \cite{KO}.
We also study non-abelian theories but the analysis is far
from complete. We only give several simple examples of
physical states for the gauge group SL(2, \R).
\par
%
%
We consider a topological gauge theory with a gauge group G
in two dimensions.
The topology of the two-dimensional surface is chosen to be
$M = \R \times {\rm S}^1$ with coordinates
$x^\mu\ (\mu = 0, 1)$, $x^0 \in \R$, $0 \leq x^1 < 2\pi$.
The action, which can be obtained from the Chern-Simons action
in three dimensions by a dimensional reduction, is given by
\be
S = {1 \over 2\pi} \int_M d^2 x \, \epsilon^{\mu\nu}
\tr \left( \phi F_{\mu\nu} \right),
\label{action}
\ee
where $\phi$ is a scalar field in the adjoint representation of G
and $F_{\mu\nu} = \partial_\mu A_\nu - \partial_\nu A_\mu
+ [ A_\mu , A_\nu ]$ is the field strength of the gauge field
$A_\mu$.
These fields are matrix valued and can be expanded in generators
$T_a$ $(a = 1, \cdots, {\rm dim\; G})$ of the Lie algebra of G,
e.g. $\phi = \phi^a T_a$. The generators satisfy
\be
[ T_a , T_b ] = f_{ab}{}^c \, T_c, \qquad
\tr \left( T_a T_b \right) = {1 \over 2} \eta_{ab},
\label{trace}
\ee
where $f_{ab}{}^c$ and $\eta_{ab}$ are the structure constant and
the Killing metric of the Lie algebra of G.
Our convention for the antisymmetric tensor $\epsilon^{\mu\nu}$ is
$\epsilon^{01} = +1$.
\par
The action (\ref{action}) is invariant under the gauge
transformation
\be
\delta A_\mu = \partial_\mu \epsilon + [ A_\mu , \epsilon ], \qquad
\delta \phi = [ \phi , \epsilon ].
\label{gaugetrans}
\ee
We use the light-cone gauge \cite{GK} $A^a_- = 0$ to fix the gauge
invariance, where we have defined the light-cone directions
$x^\pm = {1 \over \sqrt{2}} (x^0 \pm x^1)$.
Following the standard Faddeev-Popov procedure
we obtain the gauge fixed action
\be
S_{\rm GF} = -{1 \over \pi} \int_M d^2 x \tr \left(
A_+ \partial_- \phi + b_+ \partial_- c \right),
\label{gfaction}
\ee
where $b_+, c$ are the Faddeev-Popov ghost fields.
It should be noted that the bosonic fields $A_+, \phi$ and the
fermionic fields $b_+, c$ have the same spin content and
the same first order action.
The gauge fixed action (\ref{gfaction}) is invariant under the
nilpotent BRST transformation
\ba
\delta_B A_+ \A = \A \partial_+ c + [ A_+, c ], \nonu
\delta_B \phi \A = \A [ \phi , c ], \nonu
\delta_B b_+ \A = \A - \partial_+ \phi - [ A_+, \phi ]
- \{ b_+ , c \}, \nonu
\delta_B c \A = \A - c^2.
\label{brsttrans}
\ea
The BRST charge is obtained by the Noether procedure as
\be
Q_B = -{1 \over \sqrt{2} \pi} \int d x^1 \, \tr
\left[ c \left ( \partial_+ \phi
+ [ A_+, \phi ] \right) + b_+ c^2 \right].
\label{brstcharge}
\ee
\par
Quantization of field theories with first order actions such as
Eq.\ (\ref{gfaction}) was generally discussed in Ref.\ \cite{FMS}.
Solutions of the equations of motion are expanded in oscillator
modes as
\ba
A_+^a \A = \A {1 \over \sqrt{2}\, i} \sum_{n\in\Z}
\beta_n^a \, z^{-n},
\qquad \phi_a = \sum_{n\in\Z} \gamma_{na} \, z^{-n}, \nonu
b_+^a \A = \A {1 \over \sqrt{2}\, i} \sum_{n\in\Z}
b_n^a \, z^{-n}, \qquad\; c_a = \sum_{n\in\Z} c_{na} \, z^{-n},
\label{modeexp}
\ea
where $z = \e^{\sqrt{2} i x^+}$.
The coefficients satisfy the canonical (anti-)commutation relations
\ba
\A\A [ \gamma_{ma}, \beta_n^b ] = \delta_a^b \delta_{m+n, 0}, \qquad
\{ c_{ma}, b_n^b \} = \delta_a^b \delta_{m+n, 0}, \nonu
\A\A \qquad \mbox{other (anti-)commutators} = 0
\label{ccr}
\ea
and the hermiticity conditions
\ba
\beta^{a \dagger}_n \A = \A - \beta^a_{-n}, \qquad
\gamma^\dagger_{na} = \gamma_{-na}, \nonu
b^{a \dagger}_n \A = \A b^a_{-n}, \qquad\;\;\;\,
c^\dagger_{na} = c_{-na}.
\label{hermiticity}
\ea
\par
Following Ref.\ \cite{FMS} we introduce the $q$-vacua $\ket{q}_B$
and $\ket{q}_F$ for the $\beta$-$\gamma$ and $b$-$c$ systems
respectively. They are defined by
\ba
\beta^a_n \ket{q}_B \A = \A 0 \quad ( n \geq -q_a ), \qquad
\gamma_{na} \ket{q}_B = 0 \quad ( n \geq q_a+1 ), \nonu
b^a_n \ket{q}_F \A = \A 0 \quad ( n \geq -q_a ), \qquad\,
c_{na} \ket{q}_F = 0 \quad ( n \geq q_a+1 ),
\label{qvacuum}
\ea
where $q$ denotes a set of integers
$q_a\ (a = 1, \cdots, {\rm dim\; G})$. The ordinary SL(2, \C)
invariant vacua are $\ket{q=0}_B$ and $\ket{q=0}_F$.
The Fock spaces for the $\beta$-$\gamma$ and $b$-$c$ systems are
constructed on these $q$-vacua.
For the bosonic $\beta$-$\gamma$ system, $q$-vacua with different
values of $q$ generate different Fock spaces.
On the other hand, all $q$-vacua for the fermionic $b$-$c$ system
generate the same Fock space. Indeed, the general fermionic
$q$-vacuum can be obtained from the SL(2, \C) invariant vacuum
by acting the oscillator modes as
\be
\ket{q}_F = \prod_a \ket{q_a}_a, \qquad
\ket{q}_a = \left\{ \matrix{
\hfill b^a_{-q} b^a_{-q+1} \cdots b^a_{-1} \ket{0}_a
\qquad ( q > 0 ), \cr
\hfill c_{q+1 a} c_{q+2 a} \cdots c_{0a} \ket{0}_a
\qquad ( q < 0 ). \cr} \right.
\label{fqvacuum}
\ee
The $q$-vacuum of the total system is defined by
\be
\ket{q} \equiv \ket{q}_B \otimes \ket{-q-1}_F.
\label{totalqvacuum}
\ee
A choice of the vacuum in the fermionic sector is a matter of
conventions since all the $q$-vacua generate the same Fock space as
discussed above. The choice in Eq.\ (\ref{totalqvacuum}) is
convenient for later analyses. If the ghost number of $\ket{0}$ is
defined to be zero, the ghost number of $\ket{q}$ is $\sum_a q_a$.
With respective to the $q$-vacuum (\ref{totalqvacuum}) the modes
$\beta^a_n, c_{an}\ (n \geq -q_a)$ and
$\gamma_{na}, b^a_n\ (n \geq q_a+1)$ are annihilation operators,
while other modes are creation operators.
The whole space of states is $\bigoplus_q F_q$, where
$F_q$ is the Fock space constructed on
the $q$-vacuum (\ref{totalqvacuum}).
\par
The bra $q$-vacua conjugate to the above ket $q$-vacua
${}_B \bra{q} \equiv (\ket{q}_B)^\dagger$,
${}_F \bra{q} \equiv (\ket{q}_F)^\dagger$ satisfy
\ba
{}_B \bra{q} \beta_n^a \A = \A 0 \quad ( n \leq q_a ), \qquad
{}_B \bra{q} \gamma_{na} = 0 \quad ( n \leq -q_a-1 ), \nonu
{}_F \bra{q} b^a_n \A = \A 0 \quad ( n \leq q_a ), \qquad\,
{}_F \bra{q} c_{na} = 0 \quad ( n \leq -q_a-1 ).
\label{braqvacuum}
\ea
The bra $q$-vacuum of the total system is defined by
\be
\bra{q} \equiv {}_B \bra{q} \otimes {}_F \bra{-q-1}
\label{totalbraqvacuum}
\ee
and satisfies the orthonormality condition
\be
\langle -q-1 \vert q' \rangle = \delta_{q, q'}.
\label{norm}
\ee
\par
In terms of the oscillator modes the BRST charge is given by
\be
Q_B = \sum_{n \in \Z} c_{-n\, a} \left( J^a_{{\rm matt}\, n}
            + {1 \over 2} J^a_{{\rm gh}\, n} \right),
\label{charge}
\ee
where the gauge currents for $\phi$, $A_\mu$ and the ghost fields are
\be
J^a_{{\rm matt}\, n} = i n \gamma^a_n
- i f_{bc}{}^a \sum_{m\in\Z} \beta^b_{n-m} \gamma^c_m, \qquad
J_{{\rm gh}\, n}^a = - i f_{bc}{}^a \sum_{m\in\Z} b^b_{n-m} c^c_m.
\label{gaugecurrent}
\ee
The BRST charge can be shown to be hermitian $Q_B^\dagger = Q_B$
and nilpotent $Q_B^2 = 0$. The total gauge current $J_n^a$ and the
Virasoro generators $L_n$ can be expressed as BRST-exact forms
\ba
J_n^a \A = \A J^a_{{\rm matt}\, n} + J^a_{{\rm gh}\, n}
= \{ Q_B, b^a_n \}, \nonu
L_n \A = \A \sum_{m\in\Z} m \, N_q \! \left( \beta^a_{n-m}
\gamma_{ma} + b^a_{n-m} c_{ma} \right) = \{ Q_B, G_n \},
\label{brstexact}
\ea
where $N_q$ is the normal ordering with respect to the
$q$-vacuum (\ref{totalqvacuum}) and
$G_n = \sum_{m\in\Z} b^a_{n-m} \beta_{m \, a}$.
\par
%
%
Now we shall obtain physical states for abelian theories,
{\it i.e.}, G = SO(2) or SO(1, 1).
Physical states are determined from the BRST cohomology.
For the abelian theory the BRST charge has a simple form
\be
Q_B = i \sum_{n\in\Z} n c_{-n} \gamma_n.
\label{abelbrst}
\ee
By definition the physical states satisfy $Q_B \ket{\psi} = 0$ and
two states which differ by a BRST-exact state are identified:
$\ket{\psi} \sim \ket{\psi} + Q_B \ket{\chi}$.
\par
Using Eqs.\ (\ref{qvacuum}), (\ref{braqvacuum}) it can be shown
that the total $q$-vacua (\ref{totalqvacuum}),
(\ref{totalbraqvacuum}) are BRST invariant
$Q_B \ket{q} = 0 = \bra{q} Q_B$ for all $q \in \Z$.
The BRST transformations of the oscillator modes are
\ba
[ Q_B , \beta_{-n} ] \A=\A i n c_{-n}, \qquad\;\,
\{ Q_B , c_{-n} \} = 0,  \nonu
\{ Q_B , b_{-n} \} \A=\A - i n \gamma_{-n}, \qquad
[ Q_B , \gamma_{-n} ] = 0.
\label{brstc}
\ea
Therefore, a set of the nonzero modes
$( \beta_{-n}, c_{-n}, b_{-n}, \gamma_{-n} )\ (n \not= 0)$
forms the BRST quartet \cite{KO}, and the zero modes
$\beta_0, c_0, b_0, \gamma_0$ are BRST singlets.
It should be noted that creation operators are transformed to
creation operators while annihilation operators are transformed to
annihilation operators.
This is due to the particular choice of the fermionic vacuum in
Eq.\ (\ref{totalqvacuum}). If we used a different fermionic
$q$-vacuum in Eq.\ (\ref{totalqvacuum}), creation operators and
annihilation operators were mixed under the BRST transformation.
\par
Following Ref.\ \cite{KO} we introduce the projection operator
$P^{(0)}$ onto the subspace generated by only the zero modes
in the Fock space $F_q$
\ba
P^{(0)} = \left\{ \matrix{ \hfill \displaystyle \sum_{m=0}^\infty
\sum_{n=0}^1 {1 \over m!} ( \gamma_0 )^m
( b_0 )^n \ket{q} \bra{-q-1} ( c_0 )^n
( \beta_0 )^m \quad ( q \geq 0 ), \cr
\hfill \displaystyle \sum_{m=0}^\infty
\sum_{n=0}^1 {1 \over m!} ( \beta_0 )^m
( c_0 )^n \ket{q} \bra{-q-1} ( b_0 )^n
( \gamma_0 )^m \quad ( q < 0 ),}\right.
\label{zeroproj}
\ea
which commutes with the BRST charge. The projection
operators $P^{(k)}$ onto the $k$ nonzero mode sector for $k \geq 1$
are expressed recursively as
\ba
P^{(k)} \A=\A {1 \over k} \sum_{n=-q}^\infty \left( - \gamma_{-n}
P^{(k-1)} \beta_n + b_{-n} P^{(k-1)} c_n \right)  \nonu
\A \A + {1 \over k} \sum_{n=q+1}^\infty \left( \beta_{-n}
P^{(k-1)} \gamma_n + c_{-n} P^{(k-1)} b_n \right),
\label{kproj}
\ea
which also commute with the BRST charge.
These projection operators are orthogonal, complete and hermitian:
\be
P^{(k)} P^{(l)} = \delta_{k,l} P^{(k)}, \qquad
\sum_{k=0}^\infty P^{(k)} = 1, \qquad
P^{(k) \, \dagger} = P^{(k)}.
\label{projection}
\ee
We can rewrite $P^{(k)}$ for $k \geq 1$ in a BRST exact form
\ba
P^{(k)} \A=\A \{ Q_B , R^{(k)} \}, \nonu
R^{(k)} \A=\A - {i \over k} \sum_{n=-q}^\infty {1 \over n} b_{-n}
P^{(k-1)} \beta_n - {i \over k} \sum_{n=q+1}^\infty {1 \over n}
\beta_{-n} P^{(k-1)} b_n.
\label{P=QR}
\ea
\par
Using the projection operators we can express the general
solution of $Q_B \ket{\psi} = 0$ as
\ba
\ket{\psi} \A=\A \sum_{k=0}^\infty P^{(k)} \ket{\psi}  \nonu
\A=\A P^{(0)} \ket{\psi} + Q_B \left( \sum_{k=1}^\infty R^{(k)}
\ket{\psi} \right).
\label{sol}
\ea
{}From this we conclude that the physical subspace consists of
states generated by only the zero modes. Therefore we have an
infinite number of physical states
\ba
( \gamma_0 )^m ( b_0 )^n \A \ket{q} \A \qquad
\mbox{for $q \geq 0$}, \nonu
( \beta_0 )^m ( c_0 )^n \A \ket{q} \A \qquad
\mbox{for $q < 0$},
\label{physical}
\ea
where $m$ is an arbitrary non-negative integer and $n$ is 0 or 1.
The ghost numbers of these states are $q-n$ and $q+n$ respectively.
Similarly the bra physical states are
\ba
\bra{-q-1} ( c_0 )^n ( \beta_0 )^m \A \A \qquad
\mbox{for $q \geq 0$}, \nonu
\bra{-q-1} ( b_0 )^n ( \gamma_0 )^m  \A \A \qquad
\mbox{\rm for $q < 0$},
\label{braphysical}
\ea
where $m, n$ take the same values as in Eq.\ (\ref{physical}).
Since each state in Eq.\ (\ref{physical}) has a nonvanishing
product with one of the states in Eq.\ (\ref{braphysical}),
the physical states in
Eqs.\ (\ref{physical}), (\ref{braphysical}) are not BRST-exact.
\par
%
%
Next we shall consider the BRST cohomology of the G = SL(2, \R)
theory \cite{FK,TERAO} as an example of non-abelian cases.
The generators of the SL(2, \R) Lie algebra satisfy
\be
[ T_3, T_\pm ] = \pm T_\pm, \qquad [ T_+, T_- ] = T_3
\label{slalg}
\ee
and non-zero components of the Killing metric are
$\eta_{+-} = \eta_{-+} = \eta_{33} = 1$. To find physical
states we can consider, without loss of generality,
eigenstates of the Virasoro generator $L_0$ and the charge operator
in the Cartan subalgebra $J^3_0$, which commute
with the BRST charge and commute each other.
Since these operators are BRST-exact (\ref{brstexact}), non-trivial
physical states must have zero eigenvalue for both of them.
\par
Let us find the condition for which the $q$-vacuum is a non-trivial
physical state. The $q$-vacuum defined by
Eqs.\ (\ref{totalqvacuum}), (\ref{qvacuum}) is
an eigenstate of $L_0$ and $J^3_0$
\be
L_0 \ket{q} = 0, \qquad J^3_0 \ket{q} = 2 i (q_+ - q_-) \ket{q}.
\label{eigen}
\ee
Then, the $q$-vacua with zero eigenvalue for $L_0$ and $J_0^3$ and
satisfying $Q_B \ket{q} = 0$ are
\be
\ket{q}: \quad -1 \leq q_3 \leq 2 q_+ + 1, \quad q_+ = q_-.
\label{invcond}
\ee
On the other hand, the bra $q$-vacua with zero eigenvalue for
$L_0$ and $J_0^3$ and satisfying $\bra{-q-1} Q_B = 0$ are
\be
\bra{-q-1}: \quad 2 q_+ \leq q_3 \leq 0, \quad q_+ = q_-.
\label{brainvcond}
\ee
Some of the states in Eqs.\ (\ref{invcond}), (\ref{brainvcond}) may
be BRST-exact and therefore orthogonal to all physical states.
The ket state is not BRST-exact if it has a nonvanishing product
with at least one physical bra state in Eq.\ (\ref{brainvcond}).
There are only two such states in Eq.\ (\ref{invcond}):
$\ket{q}\ (q_+ = q_- = q_3 = 0,\; -1)$.
The bra states which have nonvanishing products with these two
states are $\bra{-q-1}\ (q_+ = q_- = q_3 = 0,\; -1)$.
\par
There are also physical states which contain the oscillator modes.
Simple examples of them are $(\gamma^3_0)^m (b^3_0)^n \ket{0}$,
where $m$ is an arbitrary non-negative integer and $n$ is 0 or 1.
We have not obtained all such physical states containing the
oscillator modes. It seems much more difficult to obtain the
complete BRST cohomology for non-abelian cases than the abelian
case. One origin of the difficulty is that the $q$-vacua do not
form representations of the gauge group G by themselves.
Acting the SL(2, \R) charge $J^a_0$ on a $q$-vacuum gives a state
which contain the oscillator modes in general.
To obtain all physical states it is better first to
classify all states according to representations of G.
\par
%
%

%
\end{document}